\documentclass[journal,12pt,draftclsnofoot,onecolumn]{IEEEtran}
\IEEEoverridecommandlockouts
\ifCLASSINFOpdf
\else
\fi
\usepackage[cmex10]{amsmath}
\usepackage{amsthm,cite,graphicx,,booktabs,lipsum,color,bm,algorithm,subcaption}
\usepackage{algpseudocode,pifont,tikz,paralist,multirow,xcolor}

\begin{document}
\begin{titlepage}
\begin{center}
\vspace*{-2\baselineskip}
\begin{minipage}[l]{7cm}
\flushleft
\includegraphics[width=2 in]{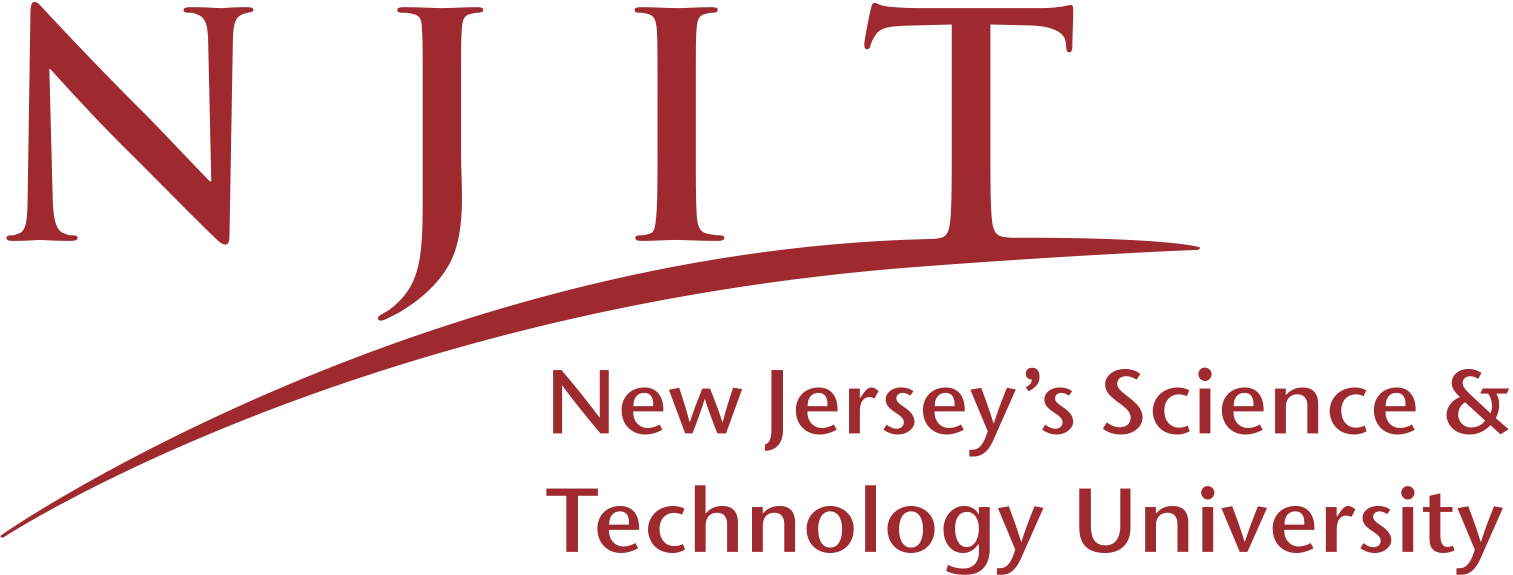}
\end{minipage}
\hfill
\begin{minipage}[r]{7cm}
\flushright
\includegraphics[width=1 in]{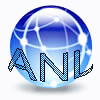}
\end{minipage}

\vfill

\textsc{\LARGE Green Cloudlet Network: A Distributed Green Mobile Cloud Network\\[12pt]}
\vfill
\textsc{
\LARGE  XIANG SUN \\ NIRWAN ANSARI}\\
\vfill
\textsc{\LARGE TR-ANL-2015-008\\[12pt]
\LARGE Dec 21, 2015}\\[1.5cm]
\vfill
{ADVANCED NETWORKING LABORATORY\\
 DEPARTMENT OF ELECTRICAL AND COMPUTER ENGINEERING\\
 NEW JERSY INSTITUTE OF TECHNOLOGY}
\end{center}
\end{titlepage}
\title{Green Cloudlet Network: A Distributed Green Mobile Cloud Network}

\author{Xiang~Sun,~\IEEEmembership{Student~Member,~IEEE,}
        Nirwan~Ansari,~\IEEEmembership{Fellow,~IEEE,} 
\thanks{X. Sun and N. Ansari are with Advanced Networking Lab., Department of Electrical $\&$ Computer Engineering, New Jersey Institute of Technology, Newark, NJ 07102, USA. E-mail:$\{$xs47, nirwan.ansari$\}$@njit.edu.}
}


\maketitle

\begin{abstract}
This article introduces a Green Cloudlet Network (GCN) architecture in the context of mobile cloud computing. The proposed architecture is aimed at providing seamless and low End-to-End (E2E) delay between a User Equipment (UE) and its Avatar (its software clone) in the cloudlets to facilitate the application workloads offloading process. Furthermore, Software Define Networking (SDN) based core network is introduced in the GCN architecture by replacing the traditional Evolved Packet Core (EPC) in the LTE network in order to provide efficient communications connections between different end points. Cloudlet Network File System (CNFS) is designed based on the proposed architecture in order to protect Avatars' dataset against hardware failure and improve the Avatars' performance in terms of data access latency. Moreover, green energy supplement is proposed in the architecture in order to reduce the extra Operational Expenditure (OPEX) and $CO2$ footprint incurred by running the distributed cloudlets. Owing to the temporal and spatial dynamics of both the green energy generation and energy demands of Green Cloudlet Systems (GCSs), designing an optimal green energy management strategy based on the characteristics of the green energy generation and the energy demands of eNBs and cloudlets to minimize the on-grid energy consumption is critical to the cloudlet provider.
\end{abstract}

\IEEEpeerreviewmaketitle

\section{Introduction}
As our mobile phones and tablets are getting much smarter, a big shift of user preference from traditional desktops and laptops to smart phones and tablets is merging, as indicated in the 2014--2019 global mobile data traffic forecast from Cisco: there were almost 7.4 billion global mobile devices and connections in use in 2014 and will grow to 11.5 billion by 2019 \cite{1}. Meanwhile, an increasing number of intelligent mobile applications have attracted more people to use smart portable devices, which consume more mobile energy and generate more traffic. However, some computing intensive applications, such as speech recognition, image processing, video analysis, online games and augmented reality, are impracticably implemented in portable devices due to the resource limitation. The emergence of Mobile Cloud Computing (MCC) technology alleviates the challenge of resource constraint and battery life shortage of portable devices by offloading some computing and communications intensive application workloads into the cloud.

The existing MCC platforms are all cloud-based architecture \cite{1.1}. Specifically, smart User Equipments (UEs) transmit their application workloads to the cloud via Wide Area Network (WAN). A bunch of VMs in the cloud assist UEs running their offloaded application workloads, and UEs only need to do some simple operations, such as sensing the environment, issuing orders to VMs, etc. However, the communications links between VMs and UEs, which traverse WAN, may incur long End-to-End (E2E) delay \cite{2}. Meanwhile, the E2E delay is critical for the MCC applications. It is reported that augmented reality applications require an E2E delay of less than 16 $ms$ \cite{2.1} and the cloud-based virtual desktop applications require an E2E delay of less than 60 $ms$ \cite{2.2}. Thus, the long E2E delay of the interaction between UEs and VMs via WAN deters the usability of MCC applications. Therefore, the recently proposed MCC architecture are not suitable for implement some latency intensive MCC applications.

\begin{figure}[!htb]
	\centering	
	\includegraphics[width=0.7\columnwidth]{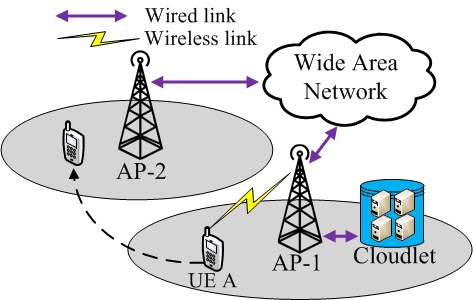}
	\caption{The pre-copy live migration procedure.}	
	\label{fig: cloudlet-based_framework}
\end{figure}

The concept of cloudlets has been proposed to eliminate the E2E delay produced in WAN. A cloudlet is a trusted, resource-rich computer or cluster of computers that is well-connected to the Internet and available for use by nearby UEs \cite{2}. Specifically, a cloudlet is a tiny version of the data center, which is deployed nearby UEs, and so UEs can access the computing resources in the nearby cloudlet through one-hop high-speed wireless local area network (e.g., LTE or WiFi). Specifically, as shown in Fig. \ref{fig: cloudlet-based_framework}, a cloudlet is connected to the wireless access point, namely, AP-1, and so UE A, which is in the coverage area of AP-1, can offload its application workloads to the VMs in the cloudlet through one wireless hop communications.

The cloudlet-based MCC framework facilitates the offloading process for UEs; however, challenges still exist. First, the low E2E delay between UEs and VMs may not be maintained when UEs roam away. For instance, as shown in Fig. \ref{fig: cloudlet-based_framework}, if the UE roams away into the coverage area of AP-2, which is not equipped with a cloudlet, the communications link between the UE and the VM should traverse WAN, and may still incur the long E2E delay. Second, although cloudlets reduce the latency between UEs and VMs, the OPEX of the cloudlet provider and $CO2$ footprint increase accordingly, i.e., extra energy is consumed for running distributed cloudlets in network.

To address the above issues, we propose a new MCC architecture, i.e., Green Cloudlet Network (GCN). The rest of this article is organized as follows. In Section II, we introduce the new architecture of GCN and some of its important components. In Section III, we unveil the challenges for designing an optimal green energy management strategy in GCN in order to minimize the OPEX of the cloudlet provider and CO2 footprint. We conclude the paper in Section IV.

\section{Architectures and vision}
\begin{figure*}[!h]
\begin{minipage}[t]{1\linewidth}
	\centering	
	\includegraphics[width=0.6\columnwidth]{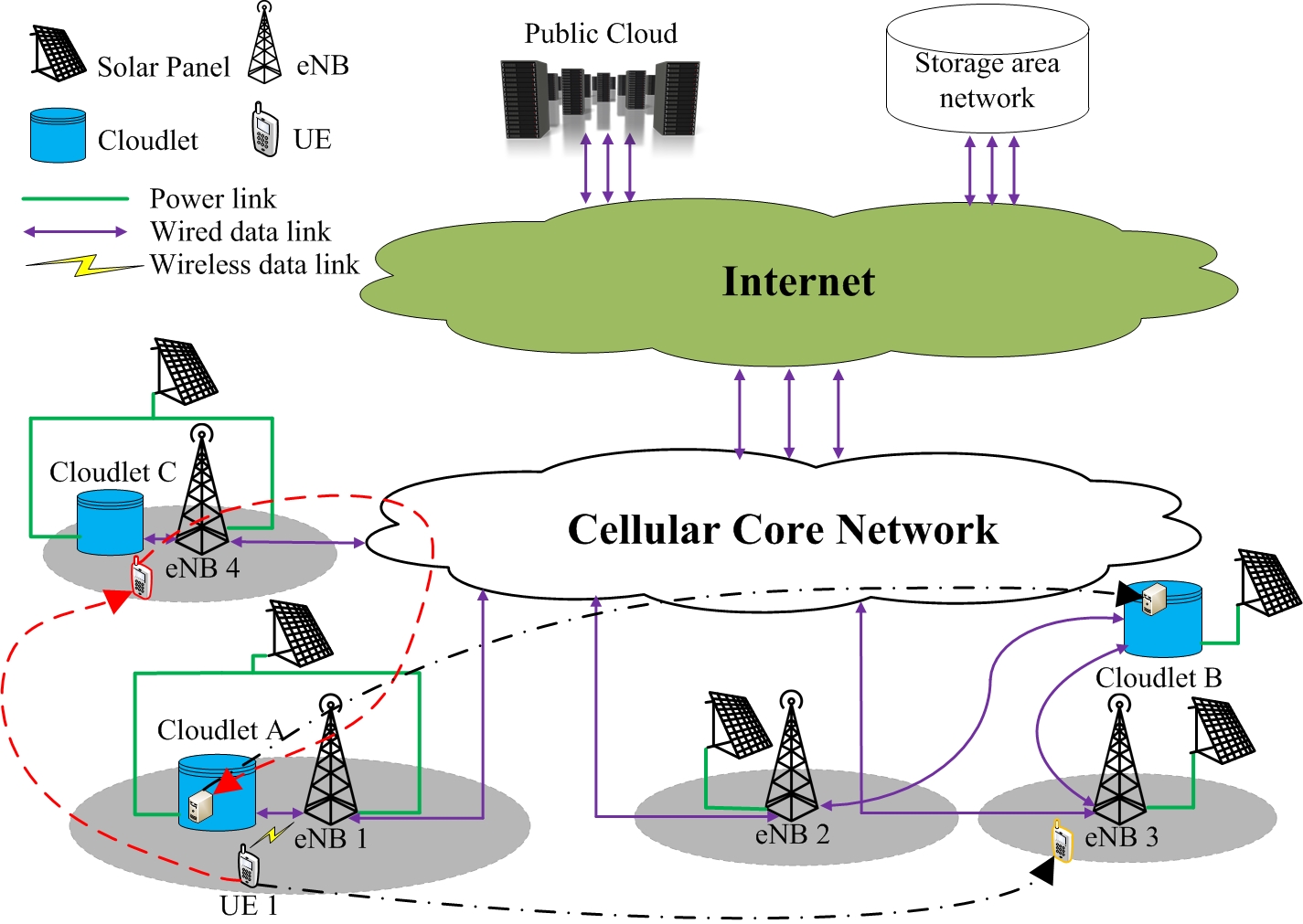}
	\caption{Green cloudlet network architecture.}	
	\label{fig:GCN_architect}
\end{minipage}
\end{figure*}

First, we introduce the concept of the Avatar, which is a software clone of the UE. Specifically, the Avatar is a VM running on the same operation system as its UE, and so the applications running in the UE can be compatibly offloaded to its Avatar. Each UE has a dedicated Avatar hosted by the cloudlet. Second, we provide an overview of the GCN architecture shown in Fig. \ref{fig:GCN_architect}, which defines how a UE connects to its Avatar, how to set up an efficient communications path between two end points (which includes UEs and Avatars in the cloudlets), etc.

\subsection{Green Cloudlet Network Architecture}
GCN is designed to provide a ubiquitous, sustainable, highly available, resilient and efficient MCC platform for UEs. By capitalizing wide distribution of eNBs in LTE networks to provide seamless connection between UEs and eNBs, a cloudlet is deployed adjacent to each eNB in the GCN architecture so that a UE's application workloads can be quickly and seamlessly offloaded from the UE to its Avatar in the cloudlet via eNB. Avatars are not only powerful computational units but also communications caches and large storage disks for their UEs. The connection between an eNB and a cloudlet can be a dedicated connection, such as high-speed fiber, so that the E2E delay between the eNB and Avatars in the cloudlet is negligible. Meanwhile, in order to reduce operational costs of running cloudlets and $CO2$ footprint, each cloudlet and eNB are powered by both on-grid energy and green energy, such as sustainable biofuels, solar and wind energy (here, we use the solar energy as an example). Moreover, every Avatar in the cloudlet can communicate with a public data center (e.g., Amazon EC2) and Storage Area Network (SAN) via the Internet in order to provision availability and reliability of the proposed architecture, i.e., if the cloudlets cannot hold UEs' Avatars anymore due to the limited capacity of the cloudlets in the network, Avatars can be migrated to the public data center to continue serving UEs, while the replicas of an Avatar's virtual disk can be stored in SAN in order to prevent data loss in case of disasters.

\begin{figure}[!htb]
	\centering	
	\includegraphics[width=1.0\columnwidth]{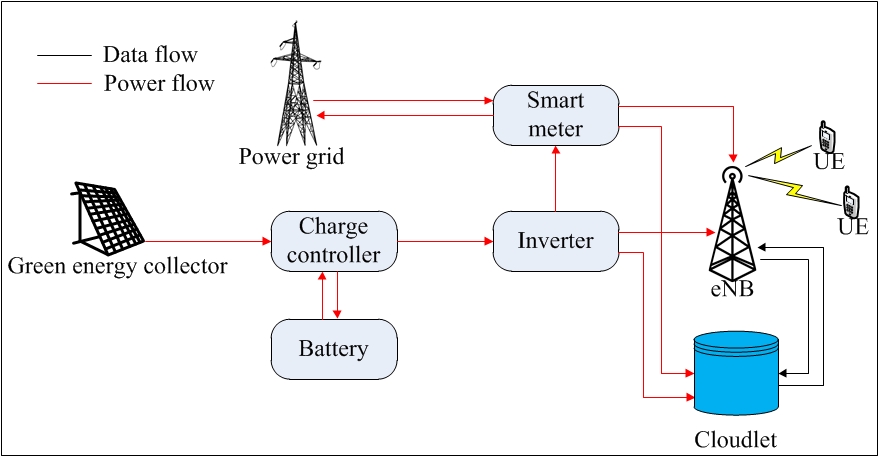}
	\caption{A green energy powered Cloudlet System.}	
	\label{fig:GCS_achitect}
\end{figure}

GCN comprises a number of geographically distributed cloudlets and eNBs connected with the cellular core network. The communications between UEs and their Avatars, Avatars and Avatars, or Avatars and the Internet should go through eNBs. The configuration of the eNB and the cloudlet can be homogeneous, i.e., one cloudlet connected with its adjacent eNB are both powered by hybrid energy. Based on the framework of green energy powered base station proposed by Han and Ansari \cite{4}, we define a Green Cloudlet System (GCS) as a basic unit as shown in Fig. \ref{fig:GCS_achitect}, in which the green energy collector extracts energy from the green energy source and converts it into electrical power, the charge controller regulates the electrical power from the green energy collector, and the inverter converts the electrical power between AC and DC. The smart meter records the electric energy from the power grid consumed by the cloudlet and eNB.

The configuration of the eNB and the cloudlet could be heterogeneous; for example, as shown in Fig. \ref{fig:GCN_architect}, if eNB 2 and eNB 3 are located in the rural area with sparse UE distribution, they can share the same Cloudlet B to provide MCC services in this area; on the other hand, picocells and femtocells are introduced in some areas with higher UE density to increase the network capacity, and so smaller size cloudlets can also be connected and shared among the cellular base stations in these areas to provide MCC services. Therefore, the cloudlet deployment strategy is still an open issue in the proposed architecture. The optimal cloudlet deployment strategy can provide sufficient but not superfluous computing and storage resources to the local UEs so that the CAPEX of the cloudlet provider is minimized and the QoS of MCC services is guaranteed. Note that the proposed GCN architecture can also facilitate the big data networking. Specifically, each UE's data streams, rather than being transmitted to the remote data center for further analysis, can be analyzed within its Avatar locally; this may significantly reduce the network delay as well as network congestion. Furthermore, the GCN architecture can benefit the Device-to-Device (D2D) communications as well \cite{4.1}. In fact, the most challenging characteristic of the D2D communications is the routing in relay by smart devices because of the mobility and the sheer number of smart devices \cite{5}. With the help of Avatars (which are statically placed in the cloudlets), the information can be shared among the smart devices through the device-Avatar-Avatar-device communications link.   

\subsection{SDN Based Cellular Core Network}
\begin{figure}[!htb]
	\centering	
	\includegraphics[width=1.0\columnwidth]{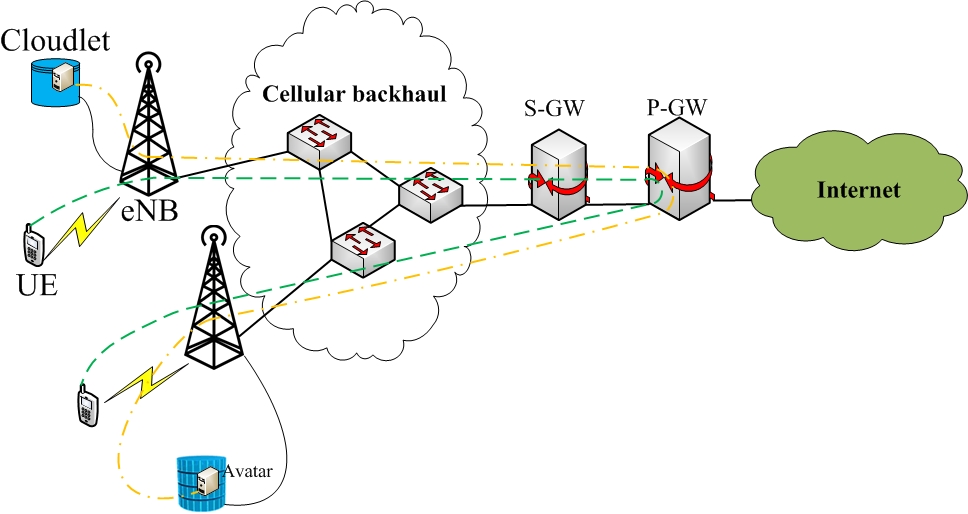}
	\caption{Communication inefficiency in the traditional EPC network.}	
	\label{fig:EPC_architect}
\end{figure}

Incorporating the distributed cloudlets into the existing mobile network burdens the traffic load of the cellular core network for two reasons. First, UEs roam among different eNBs, and thus UEs and their Avatars may not be in the same area (i.e., a UE is in the coverage area of the eNB, whose attached cloudlet does not host the UE's Avatar), that inevitably increases the traffic load of the cellular core network. For instance, as shown in Fig \ref{fig:GCN_architect}, if UE 1 roams from eNB 1's coverage area into eNB 4's coverage area and its Avatar still resides in the eNB 1's attached cloudlet, the communications path between UE 1 and its Avatar needs to traverse the cellular core network. Second, in order to keep low E2E delay, Avatars may need to be migrated from one cloudlet to another when UEs roam into a remote area\footnote{Note that the Avatar migration is triggered only when the E2E delay between a UE and its Avatar exceeds a predefined threshold.}, and so the traffic load of the cellular core network is increased because of the live Avatar migration. For instance, as shown in Fig \ref{fig:GCN_architect}, if UE 1 roams into eNB 3's coverage area and the E2E delay between the UE and its Avatar is high, which may degrade performance of the MCC applications, the Avatar should be migrated into Cloudlet B to maintain low E2E delay and the traffic, which is generated by the live Avatar migration, needs to traverse the cellular core network, thus producing extra traffic load of the cellular core network. Therefore, tremendous traffic load among cloudlets is introduced. This traffic goes through eNBs and the cellular core network without going through the Internet. Although the traditional cellular core network in terms of Evolved Packet Core (EPC) can provide guaranteed services (i.e., ensuring the E2E delay between two end points less than a threshold), it centralizes the data-plane and control-plane functionalities in the Packet data network GateWay (P-GW) and Serving GateWay (S-GW) \cite{7}. In other words, as shown in Fig. \ref{fig:EPC_architect}, all the traffic flows including D2D, Device-to-Avatar (D2A) and Avatar-to-Avatar flows (A2A) \footnote{The A2A flow includes the communications flow between two different Avatars which are associated with different UEs, and the communications flow generated by the same Avatar which has migrated from one cloudlet to another.} should go through S-GW and P-GW, thus increasing the E2E delay. Meanwhile, tremendous D2D, D2A and A2A traffic load challenges the processing capacity of S-GW and P-GW. Moreover, it is not flexible to add or change some network functionalities in EPC. Therefore, a new efficient and flexible cellular core network structure should be developed in order to support increasing traffic load in GCN.

\begin{figure}[!htb]
	\centering	
	\includegraphics[width=1.0\columnwidth]{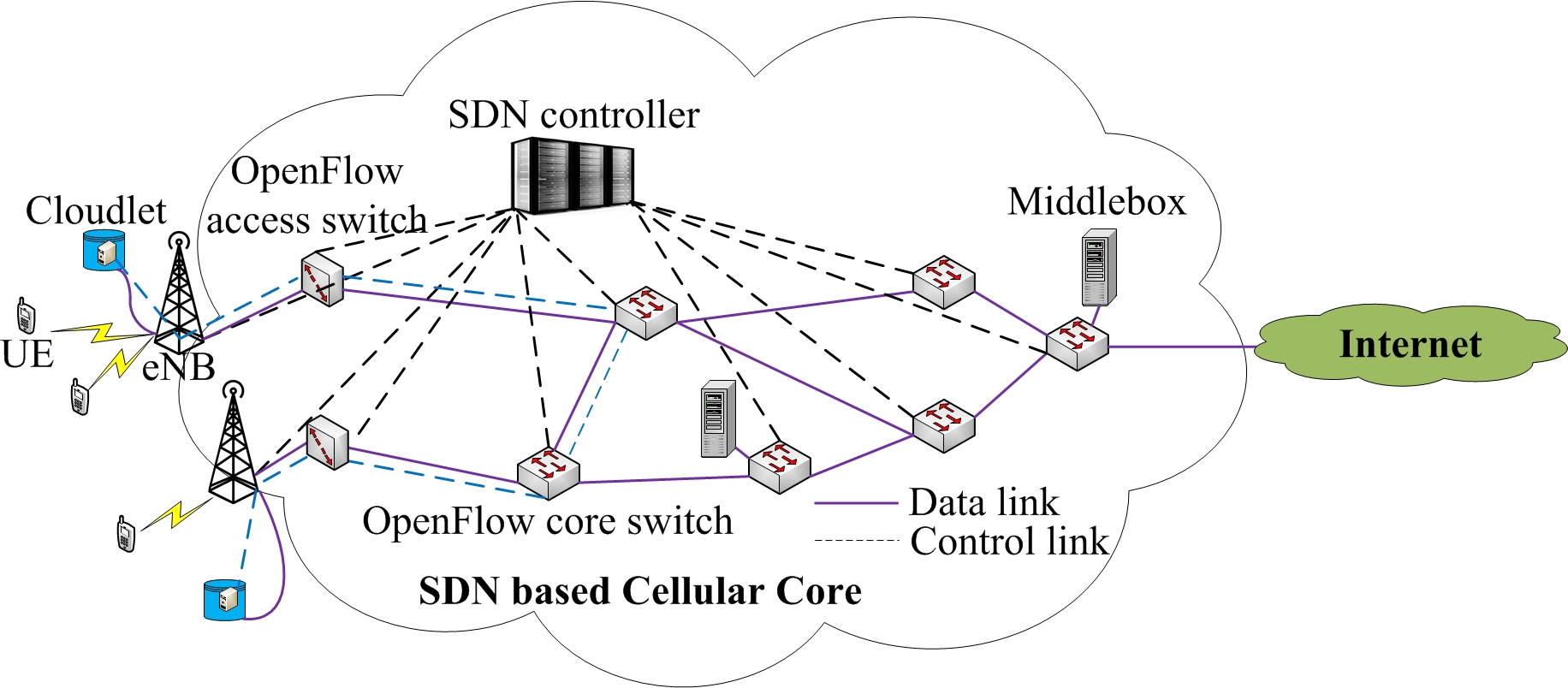}
	\caption{The SDN based cellular core network.}	
	\label{fig:SDN_architect}
\end{figure}

Applying the Software Defined Networking (SDN) technology to the cellular core network is one solution to enable a flexible and efficient network \cite{8,8.1}. The SDN architecture separates the control plane and data plane. The structure of the SDN based cellular core, as shown in Fig. \ref{fig:SDN_architect}, merges the cellular backhaul and core network together. The SDN based cellular core comprises OpenFlow switches, middleboxes (which are the appliances that the network providers can expand extra functionalities, such as network address translation, transcoder and firewall, in the network to meet various application demands), and one central controller. The SDN controller, which is a central controller, has the global information of the cellular core network, and so it facilitates any service policy by defining layer 2/3 rule. The OpenFlow switch manages the packets based on the rules in its flow tables. The controller installs the packet processing rules into different switches by using the OpenFlow protocol, which is being standardized for the signaling between the SDN switch and the controller. The middlebox provides extra flow-based service in order to efficiently use precious resources and protect the carrier from potential attacks.

The SDN based cellular core network improves the performance over the traditional EPC for three reasons. First, the SDN controller can setup an efficient and flexible routing path between two end points. Second, the SDN controller only takes charge of the control signal (no data flow) between SDN switches, thus improving the scalability as compared to the traditional EPC with centralized data plane and control plane in P-GW. Third, the controller can easily implement any network virtualized function and provision different QoS for different flows by defining new rules and installing them into SDN switches or middleboxes.

\subsection{Cloudlet Network File System}
In GCN, each Avatar is considered as a virtual machine with abundant and flexible resources, i.e., the resource capacities of Avatars are adjustable based on the resource demands of their UEs. Avatars are hosted by the corresponding servers in the cloudlets, and each server executes Avatars' applications and attaches local storages to Avatars to provision them with virtual disks. Hardware failures are normal in GCN and result in Avatar service termination and personal data loss. Therefore, designing a resilient file system such that Avatars can be quickly recovered from hardware failures is critical in the structure. Hadoop Distributed File System (HDFS) and Google File System (GFS) provide hardware fault tolerance, but they are designed for batch processing, i.e., many VMs read/write the same big file and their goal is to maximize the throughput of data access. In Cloudlet Network File System (CNFS), normally only one Avatar has the permission to access its virtual disk and most of the applications running on the Avatar focus on achieving low latency of data access on the virtual disk rather than high throughput of data access on the virtual disk, i.e., I/O latency is very intensive in CNFS. This is because the applications running in Avatars are some latency intensive tasks offloading from the UEs rather than some large data processing tasks which are invoked in the Hadoop based data center. Therefore, guaranteeing the reliability of dataset and decreasing the latency of data access are critical in CNFS.

In order to minimize the data access latency, the whole virtual disk of Avatar should be located in the same server with its CPU and memory. Meanwhile, in order to provide reliable dataset storage, a number of replicas of the Avatar's virtual disk are generated and stored in different servers. Similar to HDFS, CNFS also consists of one NameNode and a number of DataNodes. NameNode acts as a central controller to monitor the status of DataNodes (alive or dead) and the locations of different Avatars and their replicas, but it does not need to maintain the entire network's namespace. Instead, the directories and files of Avatars are maintained by the DataNode, which is a normal server deployed in the cloudlet and hosts Avatars' virtual disks and their replicas. For a fixed period of time (such as 3 seconds), every DataNode sends heartbeats to NameNode to confirm that it functions properly. If the NameNode does not receive a heartbeat from a DataNode for a certain period (such as 5 minutes), the DataNode is considered out of service and Avatars hosted in that DataNode become unavailable. The Namenode can resume these out of service Avatars from where their replicas are located and UEs can continue to be served without data loss.

Each Avatar's virtual disk and its replicas should be synchronized for each synchronization period (such as 1 minute), and so extra traffic is generated during the synchronization process. Thus, deploying more replicas for each Avatar leads to more synchronization traffic load. And the synchronization traffic needs to go through the SDN based cellular core if the replicas are located in different cloudlets. Therefore, in order to minimize the synchronization traffic load of the SDN based cellular core, it is preferred to deploy the replicas of the Avatar within the same cloudlet. However, it is not the optimal solution because the locations of the Avatar's replicas also affect the resiliency of the Avatar, the performance of the Avatar as well as the amount of migration traffic load in the SDN based cellular core. Specifically, first, in order to improve the resiliency of the Avatar, more replicas of the Avatar should be deployed in different cloudlets to minimize the probability that all the replicas are unavailable; second, as mentioned previously, when the E2E delay between a UE and its Avatar exceeds the threshold, the Avatar should be migrated to a suitable cloudlet to maintain the E2E delay at a low level. Normally, only the Avatar's CPU states and memory are transmitted to the destination cloudlet if the destination cloudlet contains one of the replicas of the Avatar. However, if the destination cloudlet does not have the Avatar's replica, the migration process may consume longer migration time and more resources are consumed (especially the bandwidth resource) by transmitting not only the Avatar's CPU states and memory but also the high volume of the Avatar's virtual disk to the destination cloudlet, which drains resources from the Avatar for executing the application workloads from its UE during the migration process, and thus degrades the performance of the Avatar consequently. Moreover, migrating the high volume of the Avatar's virtual disk will increase the traffic load of the SDN based cellular core significantly. Therefore, in order to avoid the virtual disk migration, the Avatar's replicas should be deployed where its UE commonly visits, such as home and workplace. All in all, it is beneficial to design an optimal replica placement strategy for each Avatar to minimize the traffic load (which includes the synchronization and the migration traffic load) of the SDN based cellular core and guarantee the performance and the resiliency of the Avatar.

\section{Challenges of designing the optimal green energy management in GCN}
The GCN architecture facilitates the communications between a UE and its Avatar, but it also increases the OPEX for running a number of cloudlets, i.e., a huge amount of energy is needed to maintain the cloudlet network infrastructure. It has been proved that the OPEX can be significantly reduced in the green data centers if green energy can be fully utilized \cite{10}, i.e., less on-grid energy is needed to power the data center. Therefore, “greening” is introduced in the architecture and we assume the Green Cloudlet System (GCS) is a basic unit in the GCN architecture, i.e., each eNB is attached to a dedicated cloudlet, and both of them share the same green energy generator. 

The energy demands (the sum of the energy demand of the eNB and the cloudlet) and the green energy provisioning among different GCSs exhibit the spatial dynamics, i.e., the amount of the energy demands and the green energy provisioning of different GCSs are different in the same time slot. Thus, some GCSs, which have less energy demands and more green energy provisioned, would have excess of green energy. Conversely, some GCSs, which have more energy demand and less green energy provisioned, would pull energy from the power grid. In order to minimize the on-grid energy consumption, it is beneficial to design a novel Spatial-scale Energy Balancing (SEB) strategy by adjusting the energy demands among GCSs based on each GCS's green energy provisioning.

The energy demands and the green energy generation \footnote{Note that the green energy generation is different from the green energy provisioning in a GCS, i.e., the green energy generation is the total green energy generated by the green energy collect and the green energy provisioning is the amount of green energy allocated to the eNB and its attached cloudlet in the GCS. In other words, in each time slot, the amount of the green energy generation of a GCS equals to the amount of the green energy provisioning plus the amount of green energy stored into the battery.} of a GCS exhibit the temporal dynamics, i.e., the amount of the energy demand and the green energy generation of the GCS vary over time. By regulating the battery charging and discharging, the green energy provisioning of the GCS can be adjusted over time. And it has been proved that more balanced energy gap (i.e., the ratio of the energy demand to the green energy provisioning of a GCS) among different time slots for each GCS incurs less on-grid energy consumption \cite{11}. Therefore, before running the SEB strategy to adjust the energy demands among GCSs, it is critical to design a Temporal-scale Energy Allocation (TEA) strategy to determine the amount of green energy provisioning of each GCS for each time slot so that the energy gap of the GCS among different time slots can be balanced. 


\subsection{Temporal-scale Energy Allocation (TEA) Strategy}
The parameter, Energy Drainage Ratio (EDR) \cite{11}, denoted as $\eta_{i,j}$, is adopted here to measure the energy gap between the green energy provisioning and the energy demand of GCS $i$ at time slot $j$, i.e., for any single GCS $i$ at time slot $j$, if the energy demand is $D_{i,j}$ and the allocated green energy is $E_{i,j}$, then ${\eta _{i,j}} = \frac{{{D_{i,j}}}}{{{E_{i,j}}}}$. Therefore, if $\eta_{i,j}>1$, GCS $i$ needs to consume on-gird power to accommodate the energy demand at time slot $j$. If $\eta_{i,j}<1$, the allocated green energy is enough to satisfy the energy demand of GCS $i$ at time slot $j$. The objective function of the TEA strategy is to minimize the standard deviation of every GCS's EDR vector during a time period $T$, i.e., $\min \sigma \left( {{\bm{\mathcal{Y}}_i}} \right)$,  where ${\bm{\mathcal{Y}}_i} = [{\eta _{i,1}},{\eta _{i,2}} \cdots {\eta _{i,T}}]$. Note that the smaller value of the standard deviation of a GCS's EDR vector indicates more balanced energy gap among different time slots, thus benefiting the SEB strategy to draw less on-grid energy consumption. 

To implement the TEA strategy, we need to predict the green energy generation and energy demand of each GCS at each time slot during time period $T$. The green energy generation (we consider solar energy as an example in the paper) can be accurately estimated by the existing mathematical models \cite{12}. On the other side, the energy demand of each GCS consists of two parts: first, the energy demand of the cloudlet which is determined by the number of awaked servers in the cloudlet and the amount of workload in each awaked server, and second, the energy demand of the eNB which is proportional to the amount of mobile traffic of eNB. The eNB's mobile traffic load can be estimated by using the eNB's historical mobile traffic statistics \cite{11}. However, it is difficult to estimate the energy demand of a cloudlet because it depends on the number of awaked servers and the amount of application workloads running in each Avatar hosted by the awaked servers, which may not follow the historical statistics. Therefore, establishing an energy demand prediction model of a cloudlet still remains a big challenge.

\subsection{Spatial-scale Energy Balancing (SEB) Strategy}
The energy gap is balanced at each time slot for each GCS during a time period $T$ by utilizing the TEA strategy. In each time slot, the energy gap can be further optimized by adjusting the energy demands among GCSs. The SEB strategy is proposed to balance the energy gap among GCSs so that the on-grid energy consumption of the whole GCN can be minimized. By balancing the energy gap among GCSs, the energy demand should be migrated from the GCS with lower residual green energy provisioning (i.e., GCS with larger value of EDR) to the GCS with higher residual green energy provisioning (i.e., GCS with smaller value of EDR). Thus, the objective function of the SEB strategy is to minimize the standard deviation of all the EDR vector for all GCSs in a specific time slot $j$, i.e., $\min \sigma \left( {{\bm{\mathcal{X}}_i}} \right)$, where ${\bm{\mathcal{X}}_i} = [{\eta _{1,j}},{\eta _{2,j}} \cdots {\eta _{N,j}}]$ and $N$ is the total number of GCSs in the network. In order to implement the SEB strategy, two methods can be adopted: 

\subsubsection{Adjust the Power of eNB's Pilot Signals}
One way to balance the energy gap among GCSs is to migrate the eNB's mobile traffic load from the GCS with larger value of EDR to the GCS with smaller value of EDR. Han and Ansari \cite{11,13} proposed to adjust the eNBs' coverage area by changing the power of the eNB's pilot signal so that the traffic load can be shifted among eNBs, i.e., eNB with more green energy can increase the pilot signal power to associate more UEs to undertake their traffic loads, and vice versa. Therefore, the energy gap among eNBs can be balanced.

\subsubsection{Live Avatar Migration}
\begin{figure*}[!h]
\begin{minipage}[t]{1\linewidth}
        \centering
        \includegraphics[width=1.0\columnwidth]{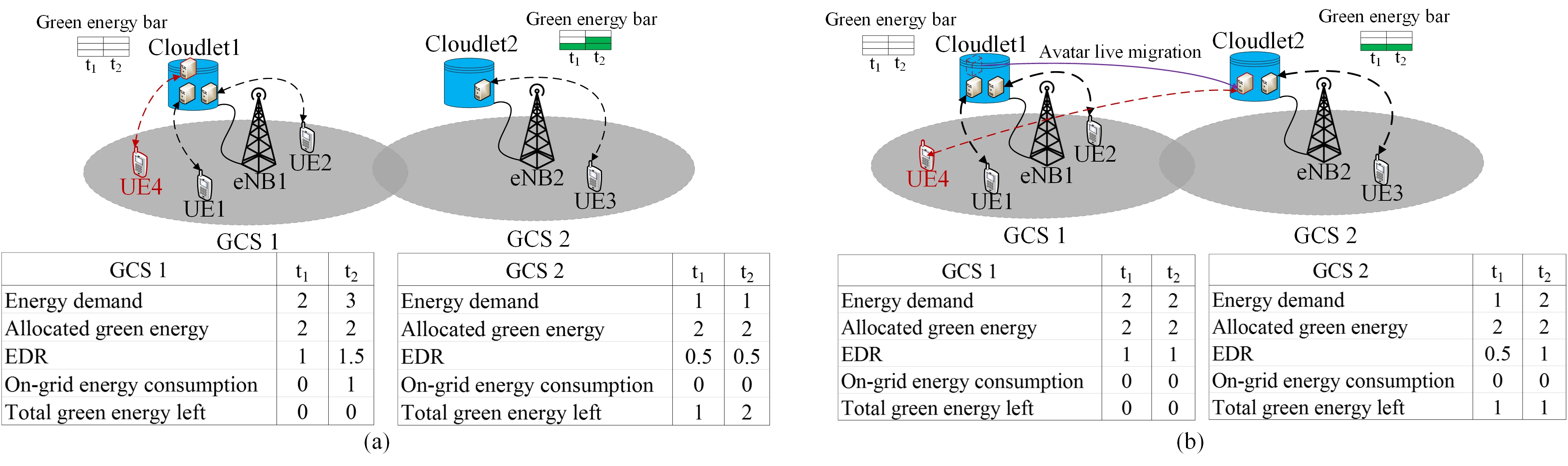}
    	\caption{Illustration of the benefit of realizing live Avatar migration in GCN.}
    	\label{fig:Avatar_migration_a}
\end{minipage}
\end{figure*}

Balancing the energy demand is not sufficient to balance the energy gap among GCSs if the major energy consumption of GCSs is from their cloudlet components, i.e., adjusting the traffic load is not sufficient enough to fill the energy gap among GCSs. The other way to implement the SEB strategy is to migrate Avatars from the GCS with larger value of EDR to the GCS with smaller value of EDR since Avatar itself can be considered as an energy consumption unit. Fig. \ref{fig:Avatar_migration_a} illustrates the benefit of live Avatar migration. Consider the two GCSs in the network and their energy demands are different. Suppose the initial residual green energy of each GCS is zero and for each time slot both of them are allocated 2 units of green energy (which are calculated by the TEA strategy). In the first time slot, there are three UEs in the network using MCC applications: UE 1, UE 2 and UE 3, where UE 1 and UE 2 are associated with their Avatars in cloudlet 1 and UE 3 is associated with its Avatar in cloudlet 2. Each Avatar consumes 1 unit of energy for running MCC applications at every time slot (here, we do not consider energy consumption of eNB since we assume that the major energy consumption of a GCS is from its cloudlet). In the second time slot, another UE, i.e., UE 4, shows up and is associated with its Avatar in cloudlet 1. We compare the two network operation strategies: 1) with no optimization, and 2) balancing the green energy gap among GCSs by adopting live Avatar migration. For the first strategy as shown in Fig. 6(a), there is no green energy remaining for GCS 1 at $t_1$ time slot and GCS 1 needs to pull 1 unit of energy from the grid in order to satisfy its energy demand at $t_2$ time slot; meanwhile, GCS 2 has a surplus of 2 units of green energy. Therefore, the network needs to consume 1 unit of on-grid energy without any optimization and $\sigma \left( {{\bm{\mathcal{X}}_1}} \right) = 0.25$ and $\sigma \left( {{\bm{\mathcal{X}}_2}} \right) = 0.5$. In the second strategy as shown in Fig. 6(b), when UE 4 shows up, GCS 1 optimizes the energy gap by migrating UE 4's Avatar from cloudlet 1 to cloudlet 2 at $t_2$ time slot. Therefore, GCS 1 does not need to pull any energy from the grid and only 1 unit of green energy remains for GCS 2. Meanwhile, the standard deviation of GCSs for each time slot is $\sigma \left( {{\bm{\mathcal{X}}_1}} \right) = 0.25$ and $\sigma \left( {{\bm{\mathcal{X}}_2}} \right) = 0$, i.e., applying the appropriate migration strategy can minimize the energy gap among GCSs, and can thus minimize the on-grid energy consumption.

While live Avatar migration can balance the energy gap among different cloudlets so that the on-grid energy consumption can be minimized for the entire network, two constraints need to be considered in making live Avatar migration. First, the capacity of cloudlet should be considered in making live Avatar migration decision. Second, since the E2E delay between a UE and its Avatar might increase in making green energy aware live Avatar migration (for instance, there is only one-hop delay in communications between UE 4 and its Avatar in Fig. 6(a), but if the migration occurs as shown in Fig. 6(b), UE 4 needs to go through eNB 1, SDN based cellular core network, eNB 2 and finally reaches its Avatar located in cloudlet 2, thus definitely increasing the E2E delay), the proper migration strategy should guarantee the QoS of each UE in terms of the E2E delay between a UE and its Avatar.

\section{Conclusion}
This paper proposes a new architecture, GCN, in order to provision ubiquitous MCC services to UEs so that UEs can save energy and execution time in running their applications. Meanwhile, the new architecture reduces the E2E delay between a UE and its Avatar by connecting the cloudlet directly to eNB. SDN based cellular core network is introduced in the architecture to improve the communications efficiency and flexibility as compared with the traditional EPC network. CNFS is proposed in the architecture to improve the resiliency of the system and the performance of the Avatar. Moreover, in order to reduce the OPEX of the cloudlet provider and $CO2$ footprint, green energy is provisioned in the architecture. Technical challenges of designing an optimal green energy management strategy are also discoursed in the paper. In the future, in order to further minimize the OPEX of the cloudlet providers, the spatial dynamics of the electrical cost among the cloudlets (like distributed data centers \cite{14}) may also be considered as a determinant to affect the Avatar migrations among the cloudlets.

\ifCLASSOPTIONcaptionsoff
  \newpage
\fi

\end{document}